\documentclass[aps,prl,reprint,groupedaddress]{revtex4-1}

\usepackage{Macros}

\newcommand\epsPL{\eps_{\tiny{\mbox{PL}}}}
\newcommand\rPL{R_{\tiny{\mbox{PL}}}}

\newcommand\dTotal{\nu_{\tiny{\mbox{Total}}}}
\newcommand{\rlink}{S}
\renewcommand{\refeq}[1]{(\ref{#1})}

\begin{document}

%\preprint{}

\title{Synchronization versus neighborhood similarity \\ in complex networks of non-identical oscillators}



\author{Celso Freitas}
\email[]{cbnfreitas@gmail.com}

\author{Elbert Macau}
\email[]{elbert.macau@inpe.br}

\affiliation{Associate Laboratory for Computing and Applied Mathematics - LAC, National Institute for Space Research - INPE, 
12245-970, 
S\~ao Jos\'e dos Campos, SP, Brazil}

\author{Ricardo Luiz Viana}
\email[]{viana@fisica.ufpr.br}
\affiliation{Department of Physics, 
Federal University of Paran\'a - UFPR, 81531-990, Curitiba, PR, Brazil.}


\date{\today}

\begin{abstract}
Does the assignment order of a fixed collection of slightly distinct subsystems into  given communication channels influence the overall ensemble behavior? We discuss this question in the context of complex networks of non-identical interacting oscillators. Three connection configurations found in Nature are considered here: Similar, Dissimilar and Neutral patterns. These strategies correspond 
respectively to oscillators alike, distinct and indifferent relate to its neighbors. To construct such 
scenarios we define a vertex weighted graph measure, the Total Dissonance, which comprises the sum of the dissonances between all neighbor oscillators in the network. Our numerical simulations show that the more homogeneous is a network, the higher tend to be both the coupling strength required to phase-locking and the associated final phase configuration spread over the circle. On the other hand, the initial spread of partial synchronization occurs faster for Similar patterns in comparison to Dissimilar ones, while 
neutral patterns are an intermediate situation between both extremes.
\end{abstract}


\maketitle

%\blue{
%\begin{itemize}
%\item Regular networks
%\item Community networks: 
%\item Remove $\C[<]{d} = 2$
%\item Compute the power law coefficient of the graphs
%\item Compute correlation between natural frequencies and node degree
%\item Produce an example with the same correlation between node-degree and natural frequencies and different synchronization response
%\item Fig 1: Examples of SND with Regular, community and BA
%\item Fig 2: The same $\dTotal$ distribution chart
%\item Fig 3: $\eps \times \C[<]{R}$ normalize $\eps$ by $\epsPL$
%\item Fig 4: $\eps \times S$ normalize $\eps$ by $\epsPL$ (try to merge both figures)
%\item Fig 5: $\dTotal \times \epsPL$
%\item ERROR: mean-degree 2 actually is 4! Check all.
%\end{itemize}
%}
%
%\blue{
%Nodes represent books about US politics sold by the online bookseller Amazon.com. Edges represent frequent co-purchasing of books by the same buyers, as indicated by the \aspas{customers who bought this book also bought these other books} feature on Amazon.
%}

%WHY TO STUDY THE MODELS WITH COUPLING GRAPH AND NON-IDENTICAL OSCILLATOR

Some social and biological studies about multi-agents reveal that units tend to select similar peers to interact \cite{Hamm2000,Abadzi1985}. However, there are systems which behave in opposite manner, where their components  
preferentially choose to connect themselves 
to others with some distinct inner characteristics \cite{Reusch2001}. 
In fact, Nature seems to favor the
former or the later construction, which we respectively call  \emph{Similar} or \emph{Dissimilar} (neighborhood) patterns, to achieve different agendas \cite{Lozares2013}.
This article explores ideas inspired by these scenarios within the non-identical phase oscillator Kuramoto model with local mean field coupling, which is one of the main paradigms to describe collective behavior and synchronization \cite{Kuramoto1984}. 
This model is also interesting because it approximates dynamics of a large class of non-linear oscillators near limit cycle, under weak mutual interaction \cite{Pikovsky2003}. 
Besides, this is an active research field with
a number of applications from different areas \cite{Strogatz01,Arenas2008,Filatrella2008,Follmann2014}, highlighting the fundamental role that synchronization plays.

Our numerical approach is based on 
a novel vertex weighted graph measure: the \emph{total dissonance}. This quantity can be regarded as a generalization of the classical concept of dissonance, that is, the natural frequency difference
from the case of two coupled oscillators \cite{Pikovsky2003}.
So, we define Similar and Dissimilar neighborhood patterns as the assignments of non-identical oscillators into the coupling graph which yield, respectively, significantly lower and higher total dissonance values. 
Otherwise, if a pattern has no strong bias, we call it \emph{Neutral}.
Given a fixed choice of inner properties for each oscillator and a fixed coupling graph, we search for Similar and Dissimilar patterns via an optimization algorithm interchanging oscillator's positions into the graph nodes. Finally, massive numerical simulations are performed to grasp the influence of these three different neighborhood patterns over phase-synchronization quantifiers. Regular,
scale-free, random, small world 
and community networks are considered \cite{Steen2010} to provide evidences about the ubiquity of our argument.

About related material, Ref. \onlinecite{Gardenes2011} and \onlinecite{Ji2013}, respectively, explore first and second order Kuramoto Model versions, both including local correlations between oscillator's natural frequency and node-degree. They report an explosive synchronization in the first case and cascade synchronization, according to the node-degree, in the second. Our methodology introduces a diverse relationship between natural frequencies and coupling graph, as will be discussed and illustrated in the text. 

Optimization studies also laid foundation to our research.
In Ref. \onlinecite{Brede2008}, an algorithm is proposed to construct optimized networks related to a combination of local and global synchronization measures. Their objective function is computed and refined after successive numerical integrations.
Although we follow a different approach, we point out that our results also support that \aspas{the early onset of synchronization and  rapid transition to the phase-lock are conflicting demands on the network topology} \cite{Brede2008}.
Ref. \onlinecite{Zhang2014} associates a percolation process to the spread of synchronization. In addition, they consider node interchange in the graph based on a vertex weighted graph measure. However, their characterization takes into account only the phase sign of neighbor oscillator.
Even so, we also found a similar explosive synchronization.

It is a common sense that a way to achieve more homogeneous neighborhood patterns is to gather members with closer intrinsic dynamics into communities. 
Thus, articles that investigate this framework can also benefit from our findings. 
Ref. \onlinecite{Arenas2006} addresses a Kuramoto model of identical oscillators showing that, as the transient time dies out, synchronization occurs in stages matching the granular communities of the coupling graph.
Ref. \onlinecite{Maxim2013} deals with
communities of oscillators having essentially different natural frequencies. The authors of this paper discuss ways to promote or suppress synchrony on individual subgroups.
Ref. \onlinecite{Wang09} introduces a dynamic feedback control to produce intra-communities synchronization regarding communities of identical non-linear oscillator. 
Accordingly, 
Similar, Neutral and Dissimilar patterns 
can be a feasible tool to adjust synchronization properties within the context of related works.
%
%In the following, we describe our model and the metrics we use. 
%
%Then, results about phase-lock synchronization and also its emergence are included in the following two sections.
%
\section{Model and Metrics}

We consider a system of $N$ phase oscillators, whose dynamics for the $i$-th oscillator is
\begin{equation} \label{eKM}
\dot \theta_i =
\w_i
+
\Frac{\eps}{d_i} \Sum_{j=1}^N A_{ij} \sin \C{\theta_j - \theta_i },
\end{equation}
where $\w = \C{\w_1, \ldots, \w_N} \in \RR^N$ are the oscillator's \emph{natural frequencies}. We consider $\w$ with zero mean \footnote{Without loss of generality, otherwise this mean value can be absorbed into the space stated and eliminated with a coordinate change}, randomly drawn from the uniform distribution over $\C[1]{-\pi,\pi}$. 
The \emph{coupling strength} $\eps \ge 0$ is the system parameter that adjusts the intensity of attractiveness between neighbor oscillators.  The symmetrical \emph{coupling graph} is expressed by its adjacency $N \times N$ matrix $A$, so that $A_{ii} =0$; $A_{ij} = 1$, if oscillators $i,j$ are neighbors (adjacent); and $A_{ij} = 0$, otherwise. We assume connected graphs, meaning that there is a sequence of edges joining any two vertexes in the graph. Also, $d_i := \sum_{j}^N A_{ij}$
stands for the $i$-th vertex degree. 
The Laplacian matrix is defined as $L := \diag \C{d_1, \ldots, d_N} - A$ and its eigenvalues are $0 = \lambda_1 \le \lambda_2  \le \ldots \le \lambda_N$. The first non-trivial eigenvalue $\lambda_2$, the \emph{algebraic connectivity}, is greater than zero if and only if the graph is connected \cite{Godsil2001}.  

%Note this is indeed a local mean field coupling model, because if we set $\bar \theta_i = 1/d_i \sum_{j}^N A_{ij}$ and define $\bar r_i \ee^{\C{\bar \theta_i - \theta_i}\ii} = 1/d_i \sum_{j=1}^N A_{ij} \ee^{ \C{\theta_j - \theta_i}\ii}$, then Eq. \refeq{eKM} reads $ \dot \theta_i = \w_i + \eps \bar r_i \sin \C{\bar \theta_i - \theta_i}$.

On one hand, analytical results  \cite{Jadbabaie2004} guarantee convergence to a unique (modulus $2\pi$) stable phase-locked regime, where phase differences between every two oscillators becomes constant. Precisely, this convergence occurs if the coupling strength $\eps$ is large enough in comparison with $\norm{\w} \lambda_N / \lambda_2^2$, where $\norm{.}$ denotes the Euclidean norm in $\RR^N$. Because $\lambda_2$ increases when the graph diameter $D$ is decreased and 
$\lambda_N$ decreases with its maximum degree $d_{\max}$ \cite{Li1998}, phase-locking can be achieved for smaller values of $\eps$ mostly with the reduction of $D$, but also with smaller values of $\norm{\w}$ and $d_{\max}$.
On the other hand, if we consider a system with only two phase oscillators, it is well known  \cite{Pikovsky2003} that the relation between its dissonance $\nu := \w_1 - \w_2$ versus the coupling strength $\eps$ determines the synchronization regime\footnote{Arnold Tongues for instance can describe this interplay \cite{Pikovsky2003}}. 
So, inspired by this context,
we introduce the \emph{total dissonance} measure for vertex weighted graphs as
\begin{equation} \label{edTotal}
\dTotal :=  
\Frac{1}{N}
\raiz{
	\Sum_{i,j=1}^N
A_{ij} \C{\w_i - \w_j}^2
}.
\end{equation}
Since we consider symmetrical and connected coupling graphs, it is straightforward to check that $\dTotal = 0$ if and only if all oscillator are identical. If we write $\dTotal= \dTotal\C{\w}$, this measure  quantifies how far $\w$ is from a condition where all natural frequencies are identical. 
Therefore, $\dTotal$ encompasses information about the total spreading of $\w$ by summing up individual dissonances over the coupling graph edges.

The norm of the global mean field, the \emph{order parameter}, will be denoted by 
$
R \C{\theta} = \abs{1/N \sum_{i = 1}^N \ee^{\ii \theta_i }}
$.
This quantity $R$ ranges from $0$ to $1$, respectively indicating that the
ensemble gradually changes from 
null global mean field,
where all phasors $\ee^{\ii \theta_i}$ cancel out,
to full synchronization, where $\theta_1 = \ldots = \theta_N$. 
We also make use of the \emph{edge partial synchronization index} between two oscillators $i,j$,
\[
{\rlink}_{ij} = {\rlink}_{ji}
:=
\abs{
      \Lim_{\Delta t \ra \infty}
      \Frac{1}{\Delta t} 
      \Int[t_r][t_r + \Delta t]{
      \ee^{\ii \C[0]{\theta_i \C{t} - \theta_j \C{t}}}
      }{t}
      },
\]
where $t_r$ is a large enough transient time \cite{Gardenes2007}. Oscillators $i,j$ are phase-locked, that is $\theta_i \C{t} - \theta_j \C{t}$ converges to a constant value, if and only if $\rlink _{ij} = 1$. Moreover, if this index is decreased towards zero, then weaker forms of synchronization and later uncorrelated trajectories occur \cite{Freitas2013}. We average contributions of all neighbor oscillators in the network to define the \emph{partial synchronization index}  
\begin{equation}
	\rlink := \Frac{1}{\tilde E} 
	\Sum_{i,j=1}^N A_{ij} {\rlink}_{ij},
\end{equation}
where $\tilde  E := \sum_{i,j=1}^N A_{ij}$ is the quantity of directed edges in the graph. Of course, the number of undirected edges is $E := \tilde E /2$.
Thus, $\rlink=1$ means that the whole ensemble is phase-locked, 
while $\rlink \approx 0$ yields very low coherent ensemble behavior. Note that $R\C{ \theta \C{t}}$ converges to a constant value \footnote{It is straightforward to check that this constant value cannot be the unit for non-identical oscillators.} if and only if $\rlink=1$.

An Adams-Bashforth-Moulton Method for numerical integration is applied. Transient time of at least $2.10^3$ units of time was suppressed from the data, while the convergence of approximations of $\rlink$ over successive time windows of $10^3$ units of time was the criterion to interrupt the integration.
The mean value of $R \C{\theta \C{t}}$
after the transient is denoted by $\C[<]{R}$. For a given choice of parameters and initial conditions, we indicate by $\epsPL$ the smallest
critical coupling strength $\eps>0$ inducing phase-locking, i.e. $S=1$ and $R \C{\theta \C{t}}$ converges to a constant value, which we denote by $\rPL$.

%\section{Obtaining network patterns}

Several complex networks topologies \cite{Steen2010} with $N$ nodes and $E$ (undirected) edges are considered: 
4-Regular ($N$ RE),
Barab\'asi-Albert ($N$ BA),
Erd{\H o}s-R\'enyi ($N$ ER) and 
Watts-Strogatz \footnote{Watts-Strogatz networks with rewiring probability of 0.25 are considered} ($N$ WS). 
Experiments with relatively small networks with $N=50$ are performed for the sake of easy visualization. Larger ones, with $N=500$, are also addressed to illustrate graphs closer to the theoretical degree distribution \cite{Steen2010}, yet feasible to massive numerical integration. To diminish computational cost and to allow comparison among network topologies, we consider graphs with $\tilde E = 4N$ directed edges, which yields mean node-degree $\C[<]{d} = \tilde E / N = 4$. 

Besides, a real world complex network with community structure, denoted by $105$ CO, is included in the simulations: the Krebs-Amazon Political Books network \footnote{
Data set description at
http://www.orgnet.com/ and
Adjacency matrix from
http://moreno.ss.uci.edu/data.html}. This graph comprises $N=105$ nodes, $\tilde E=882$ edges and $2$ communities. 

A single choice of natural frequencies $\w \in \RR^N$ were randomly drawn, matching network sizes $N=50,105,500$.
The term \emph{Configuration} stands for a specific pair of graph and permutation of the vector $\w$.
For a fixed coupling graph,
if one obtains permutations of $\w$ such that the associated $\dTotal$ is closer to its minimum or maximum, we have Similar (S) or Dissimilar (D) patterns, respectively. On the other hand, when no optimization process is applied and $\dTotal \C{\w}$ is far from extremes values, we call it a Neutral (N) pattern.

\begin{figure}[!htb]\centering
\includegraphics[width=\linewidth]{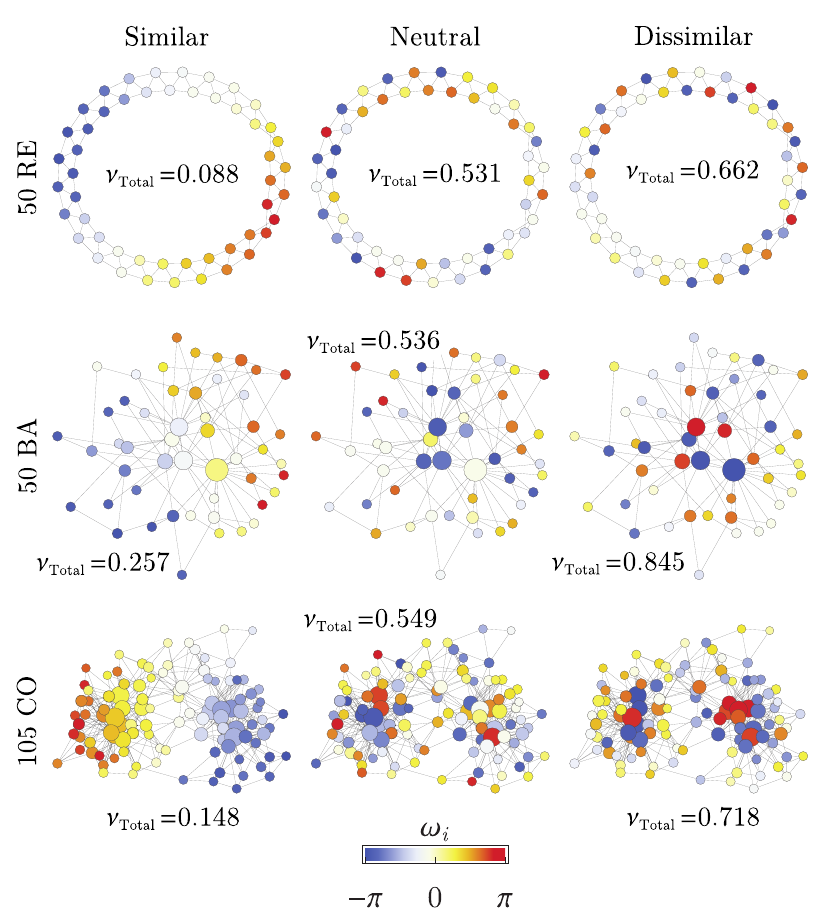}
\caption{(Color online) 
Examples of Similar, Neutral and Dissimilar patterns related 
to $N$-nodes graphs following network topology: 4-Regular ($N$ RE); Barab\'asi-Albert ($N$ BA), both with $E=100$ edges and; a community graph ($N$ CO) with $E=441$. Vertex color according to natural frequency $\w_i$ ranging from $-\pi$ (blue) to $\pi$(red); and vertex size proportional to node-degree. The associate total dissonance $\dTotal$ is displayed.
}
\label{fDtotalExamples}
\end{figure}

Fig. \ref{fDtotalExamples} illustrates graphs following RE, BA and CO topologies with Similar, Neutral and Dissimilar configurations.
From the RE graph with Similar configuration of this figure, one realizes a gradual transition of $\w_i$. It happens because this pattern reduces the dissonance through edges, so each node presents natural frequency close to the respective average of its neighbors.
However, this ordering arises differently depending on the network topology.
Hubs of the BA graph were colored with lighter tones, corresponding to the overall mean of natural frequency distribution. But in the CO graph,
positive and negative natural frequency values were placed into distinct communities, with hubs close to $\pm \pi/2$ and central nodes (in between communities) close to null $\w_i$. 

About the Dissimilar configurations from Fig. \ref{fDtotalExamples}, the opposite organization is found: each node receives natural frequency far from its neighbors. For the RE network, we clearly notice sequences of connected nodes with alternating positive and negatives values of $\w_i$. Besides, BA and CO graphs presented connected hubs with larger natural frequencies and opposite signs.
%since they are involved in more terms within the sum \refeq{edTotal} to evaluate $\dTotal$.
%
Eventually, Neutral configuration can be regarded as a blending between both previous configurations.

Let us now describe a mechanism to seek Similar and Dissimilar patterns. For a given graph, we use successive iterations of the Simulated Annealing optimization method \cite{Kirkpatrick1984} to search over the $N!$ existing permutations of the vector $\w$. This problem may be challenging by itself due to its computational complexity. 
To enhance performance, we betake the observed features described in the previous paragraph to empirically tailor our algorithm.

Let $\w^{\C[1]{1}}$ denote the input to the optimization process.
This initial seed to search Similar pattern is constructed by gathering the values of  $\C[2]{\w_1, \ldots, \w_N}$ which are close to each other into (modularity-based) communities \cite{Lancichinetti2009}. No special node ordering within each community is imposed so far, but an auxiliary step of the SA algorithm is used to reorder communities to reduce $\dTotal$.
%
%If $K$ modular-communities are detected in the graph, 
%then a partition of $\tilde \w$ consisting in $K$ chunks is done. Thus, each modular-community preliminarily receives a chunk of $\tilde \w$ with successive values of natural frequencies.
%%
%% $\tilde \w$ is divided into $K$ unbroken parts of suitable sizes assigned 
%%to the vertexes of each one, that the first community receives natural frequencies
%%
%An initial SA iteration recombines modular-communities and chunks of $\tilde \w$ to minimize $\dTotal$.
%%
For Dissimilar patterns, $\w^{\C[1]{1}}$ is allocated by matching larger vertex degrees with higher absolute values of $\C[2]{\w_1, \dots, \w_N}$. 

After that, either to minimize or maximize the total dissonance $\dTotal$, successive iterations of the SA are applied. For each stage $k=2, \ldots, 30$, its initial seed $\w^{\C[1]{k}}$ is given by the outcome of the previous iteration.
Alternating strategies to randomly evolve approximate solutions are employed at each stage. 

First, nodes to be swapped are drawn with probability proportional to vertex degree, since they are the ones which influence the most $\dTotal$. 
We also pick vertexes to be interchanged only within the same modularity-based community. Thus, we can focus the optimization process on relatively detached graph regions.
Lastly, to facilitate escape from local critical points and also to fine adjustments, we consider all nodes equiprobable.

Recall that any Configuration (of a graph $A$ and natural frequencies $\w$) which suffered no optimization process is said to be Neutral. Yet, when we apply to a Configuration $\C{A,\w}$ the optimization process described above, we obtain Similar or Dissimilar patterns with minimization or maximization $\dTotal$, respectively. So, each pair $\C{A,\w}$ generates the three patterns.

Since RE topology is deterministic and the CO graph was extracted from a data set, these classes contain a single member to be analyzed. Yet, we randomly generate and include in our experiments 100 graphs of BA, ER and WS network topologies. 

\begin{figure}
\centering
\includegraphics[scale=1]{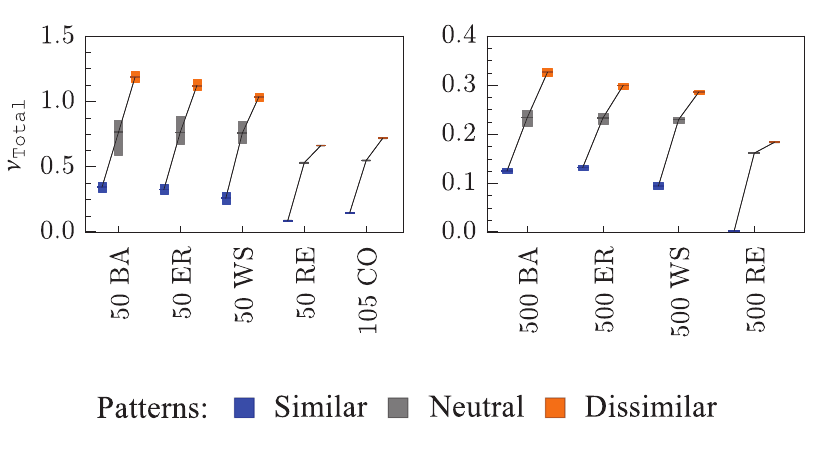}
\caption{(Color online) Total Dissonance $\dTotal$ distribution chart of different topologies with Similar (Blue), Neutral
(Gray) and Dissimilar (Orange) patterns. Data correspond to the single members from categories $50$ RE, $500$ RE and $105$ CO; and $100$ different elements for the others. 
Colored bars indicate the overall range of $\dTotal$ obtained, while mean values of each distribution are joined by a black line within each category.}
\label{fboxDTotal}
\end{figure}

Fig. \ref{fboxDTotal} displays a distribution chart of the total dissonances $\dTotal$ obtained for the categories included in this article. 
From this figure, one notices a sharp distinction among patterns since there is no $\dTotal$ range overlapping within each category.

The set of graphs with $N=500$ presented values of $\dTotal$ three times smaller than the set with $N=50$. Furthermore, both sets were qualitatively alike, which is an evidence that we were capable to produce Similar and Dissimilar neighborhoods in the large networks, at least as well as the small ones.

Because there is only a normalization by the number of vertex in the graph, but not by the quantity of edges, in Eq. \refeq{edTotal},
the total dissonance decreases with smaller mean degrees in all categories as expected. \footnote{From the optimization point of view, the normalization by any constant value, whether the number of edges $E$ or the quantity of nodes $N$, does not influence
the Similar and Dissimilar configurations obtained. We opt not to normalize by
$E$ to reflect that the more edges a graph contains, the more conflicting pairs of
dissonances exist in the network, which yields higher total dissonance.}
BA, ER and WS topologies were almost indistinguishable in the Neutral pattern, but were slightly higher in this order, for the Dissimilar case. 
All patterns of $50$ RE and $105$ CO graphs 
yielded smaller but comparable with each other total dissonance values.

We remark that the essential purpose of the total dissonance $\dTotal$ in this work was to lay foundation for the definition and search of Similar, Neutral and Dissimilar configurations. We could have formulated the subjective concepts of Similar (homogeneous) and Dissimilar (heterogeneous) networks based on variations of Eq. \refeq{edTotal}.
It is deferred for awhile to future research the study of correlations between gradual transitions of the values of this metric, far from the Similar and Dissimilar extremes, and synchronization features.

%\section{Numerical Integration}

\begin{figure*}
\includegraphics[scale=1]{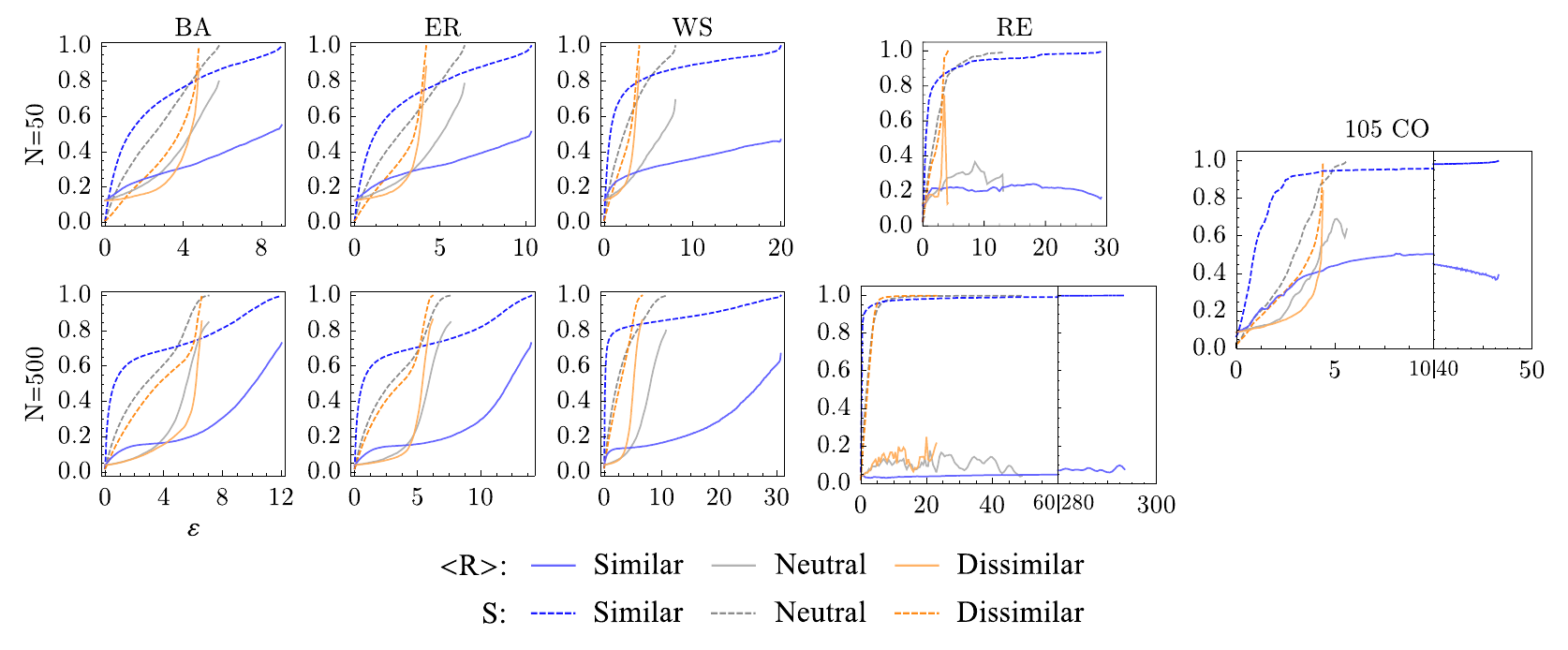}
\caption{(Color online) 
Mean order parameter after transient $\C[<]{R}$
and partial synchronization index $S$, solid and dashed lines resp.,
as a function of the coupling strength $\eps$ for different graph topologies. 
Average values of all graphs simulated within each category are shown. 
Similar, Neutral and Dissimilar cases are respectively plotted in Blue, Gray and Orange. Lines are drawn up to $\eps$ equal to the respective average critical phase-locking $\epsPL$.
}
\label{fComparaRlink}
\end{figure*}

We abuse notation and also denote by 
$S$,
$\C[<]{R}$, $\epsPL$ and $\rPL$ the associated mean values of these synchronization quantifiers considering all graphs of each category. 
A fixed random choice of initial condition $\theta^0 \in \RR^N$ is drawn from a uniform distribution over the unit circle for each network size $N$. 
So,
Fig. \ref{fComparaRlink} displays the values of $S$ and $\C[<]{R}$, as solid and dashed line resp., obtained through numerical integration for the three neighborhood configurations colored like in Fig. \ref{fboxDTotal}. The time variable of both $S$ and $\C[<]{R}$ times series are rescaled to end at the associated mean critical coupling value $\epsPL$. The $S$ lines finish at value $1$, within the numerical tolerance, while the final value of $\C[<]{R}$ lines equal to the mean critical order parameter $\rPL$.

First, one focus on the phase-locking measures $\epsPL$ and $\rPL$.
Irrespective to network size, overall results were alike. In general, 
$\epsPL$ decreases from Similar, Neutral to Dissimilar patterns; while $\rPL$ tend to increase in the same ordering. In all topologies, Similar cases demanded higher coupling strength to achieve phase-locking. In particular since this holds true even for RE networks, it shows that total dissonance patterns induce a different phenomena than the ones from Refs \onlinecite{Gardenes2011,Ji2013}.

Moreover, even when these networks phase-locked, $R$ converged to smaller values of $\rPL$. In other words, Similar ensembles tend to be harder to synchronize and to converge to regimes where oscillators were more spread around the unit circle than their counterparts.
Neutral patterns required smaller $\epsPL$ than Dissimilar ones. RE graphs were the only exception for these behavior of $\rPL$.

For all topologies, higher values of $\epsPL$ were measured when $N$ was multiplied by $10$. On the other hand, larger networks yielded
higher $\rPL$ for Similar and Neutral neighborhoods, but slightly smaller $\rPL$ for Dissimilar ones.

%In terms of differences between network topologies, BA, ER, WS graphs exhibited smaller values of $\epsPL$ in this other, if we consider Similar or Neutral patterns. Nevertheless, there is no clear bias between the coupling network topologies and $\rPL$. RE and CO graphs 

At this point, the influence of Similar, Neutral and Dissimilar patterns over the emergence of phase-synchronization is investigated, specially related to coupling strengths $\eps$ much smaller than $\epsPL$. 

Again, except for RE graphs,
we verify that Similar patterns favor weaker synchronization regimes, since the initial growth of $S$ and $\C[<]{R}$ for small coupling strength $\eps$ is more prominent. However, beyond intermediate values of $\eps$, Dissimilar patterns surpass the Similar ones through an abrupt transition. The Neutral case is in-between these two extremes, closer to the behavior of the Dissimilar group.
If we compare network topologies, BA and ER graphs displayed close values of $S$, which were smaller than WS ones for small and intermediate values of $\eps$.

%\section{Conclusion}

%Summarizing, we explored synchronization effects related to Similar, Neutral and Dissimilar neighborhood patterns, which can be also understood respectively as homogeneous, random or heterogeneous non-identical organization patterns. 

A parallel of our findings could be made with percolation of conflicting ideas, associating communication and agreement with the emergence of synchronization and phase-locking, respectively. In Similar scenarios, interaction mostly occurs among people with closely related culture backgrounds. Thus, communication can easily spread locally, but the overall population, which contains diverse members, will hardly find a compromise. 
On the other hand, when networks contain more heterogeneous neighbors, as in the Neutral and Dissimilar cases, communication demands higher effort to be established. But after that, the whole ensemble is capable to rapidly reach consensus.

In summary, experiments with several network topologies were analyzed and strong numerical trend was found. 
The Neutral case behaves in general between both extremes, closer to the Dissimilar case. 
Except for RE networks, under small coupling strength $\eps$, Similar patterns yield larger values of partial synchronization index $S$, meaning early synchronization ongoing. In contradistinction, Dissimilar ones present smaller values of $S$, but undergo abrupt increment until phase-locking. 
Moreover, all networks with Similar patterns required higher values of coupling strength to achieve phase-locking, while Dissimilar patterns converged to regimes closer to full synchronization.

\begin{acknowledgments}
We would like to thank the 
Coordena\c{c}\~ao de Aperfei\c{c}oamento
de Pessoal de N\'ivel Superior - CAPES (Process: BEX
10571/13-2), CNPq and FAPESP (grant  2011/50151-0)
for financial support. 

\end{acknowledgments}

\bibliography{References}

\end{document}